\documentclass[intlimits,twoside,a4paper]{article}

\usepackage[cp1251]{inputenc}

\usepackage[eqsecnum]{cmpj3}

\DeclareMathOperator{\Tr}{Tr}

\usepackage{amsmath}
\usepackage{bm}

\issue{2018}{21}{3}{33003}
\doinumber{10.5488/CMP.21.33003}

\title[Geometric measure of mixing of quantum state]%
{Geometric measure of mixing of quantum state}

\author[H.P. Laba, V.M. Tkachuk]{H.P. Laba \refaddr{label1},
        V.M. Tkachuk \refaddr{label2}}
\addresses{
\addr{label1} Department of Applied Physics and Nanomaterials Science,
Lviv Polytechnic National University,\\
5 Ustiyanovych St., 79013 Lviv, Ukraine\\
\addr{label2} Department for Theoretical Physics,
Ivan Franko National University of Lviv,\\
12 Drahomanov St., 79005 Lviv, Ukraine
}

\date{Received June 23, 2018}

\begin{document}

\maketitle

\begin{abstract}
We define the geometric measure of mixing of quantum state as a minimal Hilbert-Schmidt distance between the mixed state
and a set of pure states. An explicit expression for the geometric measure is obtained. It is interesting that this expression corresponds to the
squared Euclidian distance between the mixed state and the pure one in space of eigenvalues of the density matrix.
As an example, geometric measure of mixing for spin-1/2 states is calculated.

\pacs 03.65.-w,  03.67.-a
\end{abstract}

\section{Introduction}

Pure and mixed states are the key concept in quantum mechanics and in quantum information theory.
Therefore, an important question arises regarding the degree of mixing of a quantum state.
In the literature,  von Neumann entropy is often used to answer  this  question:
\begin{eqnarray}
S=- \Tr \hat\rho\ln\hat\rho=-\langle\ln\hat\rho\rangle\,,
\end{eqnarray}
which is zero for a pure state and has a maximal value for maximally mixed states. The entropy can be used as a
measure of the degree of mixing of a quantum state.
To explicitly calculate the von Neumann entropy, it is necessary to know the eigenvalue of density matrix
which is a nontrivial problem.
Therefore, the linear entropy as approximation of von Neumann entropy is also used
\begin{eqnarray}
\ln\hat\rho =\ln\left[1-(1-\hat\rho)\right]\simeq -(1-\hat\rho)\,.
\end{eqnarray}
In this approximation, the linear entropy reads
\begin{eqnarray}
S_\text{L}= \Tr\big(\hat\rho-\hat\rho^2\big)=1- \Tr\hat\rho^2.
\end{eqnarray}
Linear entropy does not satisfy the properties of von Neumann entropy.
However, to calculate the linear entropy, it is not necessary to know the eigenvalues of a density matrix.
In this case, we can directly calculate the trace of $\hat\rho^2$.
Note that $\Tr\hat\rho^2$ is called purity and is used for quantifying the degree of the purity of state.
For pure state $\hat\rho^2=\hat\rho$, and purity takes a maximal value $1$ and is less $1$ for mixed states.
A review on entropy in quantum information  can be found in book \cite{Nie10} (see also \cite{Wit18}).

Geometric ideas play an important role in quantum mechanics and in quantum information theory (for review see, for instance, \cite{Ben17}).
In our previous paper \cite{Laba17}, we use the geometric characteristics such as curvature and torsion to study the quantum evolution. The geometry
of quantum states in the evolution of a spin system was studied in \cite{Kuz16,Kuz15}. In \cite{Fry17}, the distance between quantum states was used for quantifying the entanglement of pure and mixed states.

In this paper, we use Hilbert-Schmidt distance in order to measure the  degree of mixing of quantum state.
We define the geometric measure of mixing of quantum state as minimal Hilbert-Schmidt distance between the mixed state
and a set of pure states. In section~\ref{sec2}, using this definition, we find an explicit expression for the geometric measure of mixing of quantum state.
Conclusions are presented in section~\ref{sec3}.

\section{Hilbert-Schmidt distance and degree of mixing of quantum state}\label{sec2}

To define the geometric measure of degree of mixing of quantum state, we use the Hilbert-Schmidt distance
between two mixed states. The squared Hilbert-Schmidt distance reads
\begin{eqnarray}
d^2(\hat\rho_1,\hat\rho_2)=\Tr\big(\hat\rho_1-\hat\rho_2\big)^2,
\end{eqnarray}
where $\hat\rho_1$ and $\hat\rho_1$ are density matrices of the first and the second mixed states.
We define geometric measure of mixing of quantum states as minimal squared Hilbert-Schmidt distance from the  given mixed state to a set of pure states
\begin{eqnarray}
D=\min_{|\psi\rangle}\Tr \big(\hat\rho-\hat\rho_{\textrm {pure}}\big)^2,
\end{eqnarray}
where $\hat\rho$ is density matrix of the given mixed states,
\begin{eqnarray}\label{rpure}
\hat\rho_{\textrm {pure}}=|\psi\rangle\langle\psi|
\end{eqnarray}
is density matrix of a pure state described by the state vector $|\psi\rangle$, and
minimization is done over all possible pure states.

Let us rewrite the geometric measure of mixing of quantum states as follows:
\begin{eqnarray}\label{D}
D=\min_{|\psi\rangle}\left(\Tr \hat\rho^2+\Tr \hat\rho_{\textrm {pure}}^2-2\Tr \hat\rho\hat\rho_{\textrm{ pure}}\right).
\end{eqnarray}
Three terms in (\ref{D}) can be calculated separately. For the first term, we find
\begin{eqnarray}\label{t1}
\Tr \hat\rho^2=\sum_{i}\lambda_i^2\,,
\end{eqnarray}
where $\lambda_i$ are eigenvalues of density matrix $\hat\rho$.
For pure state $\hat\rho_{\textrm {pure}}^2=\hat\rho_{\textrm {pure}}$, so the second term reads
\begin{eqnarray}\label{t2}
\Tr \hat\rho_{\textrm {pure}}^2=\Tr \hat\rho_{\textrm {pure}}=1.
\end{eqnarray}
Trace is invariant with respect to choosing the basic vectors.
To calculate the third term, we use the following
orthogonal basic vectors $|\psi\rangle,|\psi_1\rangle, |\psi_2\rangle,\ldots$ , where the first vector is equal to the state of pure state in (\ref{rpure}),  $\langle\psi|\psi_i\rangle=0$, $\langle\psi_i|\psi_j\rangle=0$, $i=1,2,\ldots$ ,  $j=1,2,\ldots$ .
Then,
\begin{eqnarray}
\hat\rho_{\textrm {pure}}|\psi\rangle=|\psi\rangle,
\end{eqnarray}
\begin{eqnarray}
\hat\rho_{\textrm {pure}}|\psi_i\rangle=|\psi\rangle\langle\psi|\psi_i\rangle=0, \ \  i=1,2,\ldots .
\end{eqnarray}
As a result, for the third term we have
\begin{eqnarray}\label{t3}
\Tr \hat\rho\hat\rho_{\textrm {pure}}=\langle\psi|\hat\rho|\psi\rangle.
\end{eqnarray}

Substituting (\ref{t1}), (\ref{t2}), (\ref{t3}) into (\ref{D}), we find
\begin{eqnarray}
D=\min_{|\psi\rangle}\left(\sum_{i}\lambda_i^2+1-2\langle\psi|\hat\rho|\psi\rangle\right).
\end{eqnarray}
This expression reaches a minimal value when $|\psi\rangle$ is equal to the eigenvector of density matrix $\hat\rho$  with maximal eigenvalue.
Thus, finally, for geometric measure of mixing of quantum state we have
\begin{eqnarray}\label{Dfin}
D=\sum_{i}\lambda_i^2+1-2\lambda_{\max}= (1-\lambda_{\max})^2+ \sum_{\lambda_i<\lambda_{\max}}\lambda_i^2.
\end{eqnarray}

For the pure state $\lambda_{\max}=1$ and all other eigenvalues are zero. Thus, for the pure state $D=0$ as it should really be.
It is interesting to note that (\ref{Dfin}) is a squared Euclidian distance in the eigenvalue space between the mixed state with eigenvalues of density
matrix $\lambda_{\max}, \ldots\lambda_i \ldots$ and pure state with eigenvalues $1,\ldots 0,\ldots$ .

One can easily find that $D$ is maximal when all eigenvalues of the density matrix are the same  $\lambda_i=1/n$, $i=1,2,\ldots n$, where $n$ is a
dimension of the quantum system. So, the maximal value of geometric measure of mixing of quantum state in this case is $D=1-1/n$
and density matrix reads
\begin{eqnarray}\label{rmax}
\hat\rho_{\max}={1\over n}\hat 1
\end{eqnarray}
and can be referred to as the maximally mixed state.

The distance between the maximally mixed (\ref{rmax}) state and the arbitrary pure (\ref{rpure}) one is
\begin{eqnarray}
d^2(\hat\rho_{\max},\hat\rho_{\textrm {pure}})=\Tr \big(\hat\rho_{\max}-\hat\rho_{\textrm {pure}}\big)^2=
\left(1-{1\over n}\right)^2+(n-1){1\over n^2}=1-{1\over n}.
\end{eqnarray}
where to calculate $\Tr$ we use the
orthogonal basic vectors $|\psi\rangle,|\psi_1\rangle, |\psi_2\rangle, \ldots ,$ where the first vector corresponds to the pure state in (\ref{rpure}).
Note that this distance is the same between the maximally mixed state and the arbitrary pure one.

At the end of this section, let us consider an explicit example of using the obtained result for calculation of geometric measure of mixing of quantum state presented by (\ref{Dfin}).
We consider the mixed state of spin-1/2 described by the density matrix
\begin{eqnarray}
\hat\rho={1\over 2}\left[1+({\bf a}\bm{\sigma})\right],
\end{eqnarray}
where $\bf a$ is Bloch vector, $\bm{\sigma}=(\sigma_x, \sigma_y,\sigma_z)$ are Pauli matrices.
Eigenvalues of this matrix are
\begin{eqnarray}
\lambda_1={1\over 2}\left(1+ a\right), \ \ \lambda_2={1\over 2}\left(1- a\right),
\end{eqnarray}
where $a=|{\bf a}|\le 1$ is the length of Bloch vector. Note that $\lambda_1$ corresponds here to $\lambda_{\max}$. Then,  according to (\ref{Dfin}), the geometric measure in this case
reads
\begin{eqnarray}
D={1\over 2}(1-a)^2.
\end{eqnarray}
At $a=1$, which corresponds to pure states (Bloch sphere) as we see $D=0$ and the mixed state is maximally mixed $D=1/2$ at $a=0$.
\section{Conclusions}\label{sec3}
We define the geometric measure of mixing of quantum state as minimal Hilbert-Schmidt distance between the given mixed state
and a set of pure states. The main problem in this definition is the procedure of minimization over pure states.
It is important that it is possible to perform this procedure  and get an explicit expression for
geometric measure of mixing of the quantum state presented by (\ref{Dfin}). This is the main result of the present paper.
It is interesting to note that (\ref{Dfin}) is the squared Euclidian distance in space of eigenvalues of the density matrix between the mixed state and the pure one.
Finally, we would like to note that similarly to the calculation of von Neumann entropy of mixed states, to calculate the
geometric measure of mixing of state, it is necessary to know the eigenvalues of the density matrix.
So, from this point of view, the difficulties of calculation of geometric
measure of mixing of states is similar to the difficulties of calculation of the
entropy measure of mixing of state.
However, definition of degree of mixing of state presented in this paper is of geometric origin and is intuitively understandable.
We hope that this result provides a new inside into the  problem under consideration.

\section*{Acknowledgements}
We thank the Members of Editorial Board for the invitation to present our results in a special issue
of Condensed Matter Physics dedicated to Prof. Stasyuk’s 80th birthday.
I (VMT) have known Prof. Stasyuk since 1978 when he delivered the lectures on Green's function method for
students of theoretical physics department. The lectures were very interesting and I thank Prof. Stasyuk for that.
We wish Prof. Stasyuk long scientific life and bright ideas in the future.

\ukrainianpart

\title{Геометрична міра змішаності квантового стану}
\author{Г.П. Лаба \refaddr{label1}, В.М. Ткачук \refaddr{label2}}
\addresses{
\addr{label1}Кафедра прикладної фізики і наноматеріалознавства, 
Національний університет "Львівська політехніка",\\
вул. Устияновича, 5, 79013 Львів, Україна
\addr{label2} Кафедра теоретичної фізики,
Львівський національний університет імені Івана Франка,\\
вул. Драгоманова, 12, 79005 Львів, Україна}
\makeukrtitle

\begin{abstract}
\tolerance=3000%
Ми означаємо геометричну міру змішаності квантоваго стану як мінімальну відстань Гільберта-Шмідта між змішаним станом та набором чистих станів. Отримано явний вираз для геометричної міри змішаності. Цікавим є те, що цей вираз відповідає квадрату евклідової відстані між змішаним та чистим станами у просторі власних значень матриці густини. Як приклад, обчислено геометричну міру змішаності станів спіна 1/2.

\keywords змішані стани, матриця густини, відстань Гільберта-Шмідта
\end{abstract}

\end{document}